\documentstyle[aps,multicol,epsfig,prb]{revtex}

\begin{document}


\title{Molecular dynamics simulations of lead clusters}
\author{S. C. Hendy}
\address{Applied Mathematics, Industrial Research Ltd, PO Box 31-310, Lower Hutt,
New Zealand}
\author{B. D. Hall}
\address{Measurement Standards Laboratory of New Zealand,
Industrial Research Ltd, PO Box 31-310, Lower Hutt,  New Zealand}
\date{\today}

\maketitle

\begin{abstract}
Molecular dynamics simulations of nanometer-sized lead clusters have been performed using the Lim,
Ong and Ercolessi glue potential (Surf. Sci. {\bf 269/270}, 1109 (1992)). The binding
energies of clusters forming crystalline (fcc), decahedron and icosahedron structures are compared,
showing that fcc cuboctahedra are the most energetically favoured of these polyhedral model
structures. However, simulations of the freezing of liquid droplets produced a characteristic form
of surface-reconstructed ``shaved'' icosahedron, in which atoms are absent at the edges and apexes 
of the polyhedron. This arrangement is energetically favoured for 600-4000 atom clusters. Larger 
clusters favour crystalline structures. Indeed, simulated freezing of a 6525-atom liquid droplet produced an
imperfect fcc Wulff particle, containing a number of parallel stacking faults. The effects of temperature 
on the preferred structure of crystalline clusters below the melting point have been considered. The 
implications of these results for the interpretation of experimental data is discussed.
\end{abstract}
\pacs{61.46.+w, 36.40.-c}

\begin{multicols}{2}
\section{Introduction}
It has long been known that atomic clusters of some fcc metals occur in stable arrangements quite
different from the bulk crystal structure. Regular non-crystalline structures, with five-fold axes 
of symmetry, were first identified
in gold \cite{ino66,Allpress67}, and are known to occur in a variety of metals \cite{Marks94}. The
stability of these arrangements relies on a high proportion of surface atoms. For example, a 4~nm
fcc lead cluster has roughly 40~\% of its atoms at the surface. The local environment of surface
atoms is quite different to those inside, and leads to distinct surface properties which play an
important role in the overall behaviour of the cluster.

Although the observation of five-fold axes of symmetry in gold clusters was unexpected, it was
immediately recognized that slightly deformed fcc tetrahedral units could be assembled to form these
structures \cite{ino66}. These alternative structures afford a net gain in binding energy for the cluster: the
strain of deforming fcc tetrahedral units, and the energy cost of twinning, is compensated for by a
higher proportion of atoms in densely packed outer faces than can occur in any faceting of a perfect fcc
particle \cite{ino69a,Marks84}.

Two characteristic types of non-crystalline structure occur: the icosahedron consists of twenty
tetrahedra assembled around a common apex; the decahedron arranges five tetrahedra around a common
edge. In each, deformation of the tetrahedra is necessary to fill the volume of the polyhedron.
This lowers the symmetry of the fcc lattice to either rhombohedral (icosahedron) or body-centered
orthorhombic (decahedron) \cite{Yang79}. In both cases, adjacent tetrahedral faces meet at a twin
plane, hence the name Multiply Twinned Particle (MTP). Packing of atoms to make a given polyhedral
form leads to sequences of atom numbers at which complete outer layers, or shells, are ``closed" (see e.g.
\cite{Martin96}). Each sequence depends on the particular polyhedral model. For example, there are
two fcc cuboctahedron sequences commonly considered: one with equilateral triangular (111) facets,
the other with hexagonal (111) facets. The closed-shell sequences for these are, respectively: 13,
55, 147, 309, 561, etc; and 38, 201, 586, 1289, etc. The hexagonal faceted cuboctahedron is often
called the ``Wulff" form, because, for noble metals, it is close to the so-called Wulff polyhedron,
which minimizes the surface energy contribution in a small fcc crystal
\cite{Herring51,Herring,Defay}. In this work we generally use the terms Wulff particle and
cuboctahedron to refer to the hexagonal and triangular forms, respectively.

Elasticity theory can explain the form and stability of MTP structures, however, more detailed
theoretical tools are needed to explore their properties: molecular dynamics (MD) has been used for
this purpose. Classical MD uses a mathematical model for the interaction potential between atoms,
and the classical equations of motion, to simulate the dynamic behaviour of an $n$-atom aggregate over
time under prescribed conditions of energy, temperature, etc. The technique is computationally
intensive, but gains in computing power continue to allow more elaborate and realistic simulations
to be performed.

Of the fcc metals, gold clusters, in particular, have been extensively studied by MD using
many-body inter-atomic potentials \cite{Ercolessi91,Lewis97,Uzi97,Uzi98,Garzon99,Soler00,Baletto00}. 
Early studies of the energetics of relaxed gold clusters suggested that crystalline structures were
favoured over MTP structures \cite{Ercolessi91}. Later investigations suggested that a
truncated Marks decahedron \cite{Marks84} may be the preferred low-temperature cluster geometry,
when there are less than about 250 atoms in the cluster, and a Wulff-particle form for larger
clusters \cite{Uzi97}. Further studies have found that disordered structures prevail for certain
small cluster sizes\cite{Garzon99}. MD has also been used to investigate the coalescence of gold 
clusters in collisions \cite{Lewis97}, structural transitions in gold clusters prior to melting 
\cite{Uzi98} and the growth of silver clusters from small seed clusters by collision \cite{Baletto00}.

The case of lead clusters has received much less attention than gold. The first comprehensive
molecular dynamics study, by Lim, Ong and Ercolessi \cite{LOE92}, utilised a many-body ``glue" potential 
to compare the energetics of certain fcc and MTP structures for a range of cluster sizes. In particular,
these authors compared the energetics of the cuboctahedron and icosahedron sequences of closed-shell structures 
(the numbers of atoms in the icosahedron sequence, conveniently, is the same as those in the cuboctahedron 
sequence). This allowed them to directly compare the
binding energies at the same numbers of atoms for each structure and they demonstrated that the fcc
crystalline structures (cuboctahedra) were favored over icosahedra for all sizes where the glue
potential is expected to hold (considered to be clusters with more than 200 atoms \cite{LOE92}).
However, subsequent simulations by these authors \cite{LOE94}, involving the quenching of a large
(8217-atom) liquid lead droplet, produced a cluster that was ``icosahedron-like". Although known not 
to be the lowest energy arrangement for that number of atoms, it was assumed to have
formed because of insufficiently long equilibration times in the simulation.

The purpose of this work is to take a more comprehensive look at lead clusters by considering a
wider range of structures and more variety in cluster sizes than in previous studies. We have also
conducted simulations of longer duration. We begin by outlining the details of our computational
procedure. This is followed by a discussion of the trends in binding energy for crystalline and
non-crystalline model structures over cluster sizes from 200 to 6000 atoms. We then consider the
melting and freezing of fcc clusters. Sections \ref{sec_resol_ico} and \ref{sec_resol_cub} then
examine the structure and energetics of clusters produced by freezing liquid droplets. Finally, we
look at the characteristic diffraction patterns produced by some MD-derived cluster structures.

\section{Computational Details}

\subsection{Molecular dynamics}

Molecular dynamics was performed using a local version of the classical molecular dynamics code
ALCMD, originally developed by Ames Laboratory. Simulations were performed in the canonical 
ensemble (i.e. constant temperature) to produce caloric curves in section IV and V.
The microcanonical ensemble (i.e. constant energy) was used to study re-solidification in section V and 
VI. The time step was chosen as $3.75 \ {\mathrm fs}$ throughout. Simulations were typically allowed to 
equilibrate for $3 \times 10^5$ time steps, or approximately one nanosecond. Longer equilibration times 
(up to $3 \times 10^6$ time steps) were used near transition points and in re-solidification studies. Following 
the equilibration period, quantities reported here (such as energies) were obtained by time-averaging over 
$1 \times 10^5$ further steps.

The inter-atomic potential used is due to Lim, Ong and Ercolessi \cite{LOE92}. This is a
many-body glue-type potential, given by
\begin{equation}
V = {1 \over 2} \sum_{ij} \phi \left( r_{ij} \right) + \sum_{i} U(n_i) \; ,
 \label{glue}
\end{equation}
where $\phi$ is a short-range two-body potential and $U$ is a many-body glue term which
reflects the effects of the conduction electrons in the metal. The $n_i=\sum_{j} \rho(r_{ij})$ is a
generalized coordination where $\rho$ is some short-ranged function. The three functions $\phi$,
$\rho$ and $U$ have been obtained by fitting to a number of known properties of lead
including cohesive energy, surface energy, elastic constants, phonon frequencies, thermal expansion 
and melting temperature \cite{Ercolessi88}. This potential has been used previously to model 
lead clusters \cite{LOE92,LOE94}, temperature-dependent surface reconstructions of low-index lead 
surfaces \cite{TOE94} and pre-melting of low-index lead surfaces \cite{Uzi95}. While the potential 
has been fitted to bulk properties of lead, the clusters studied here are sufficiently large that 
(\ref{glue}) should be considered quantitatively reliable. 

\subsection{Common neighbor analysis}

\begin{figure}
 \hspace{1.5cm}
 \epsfig{file=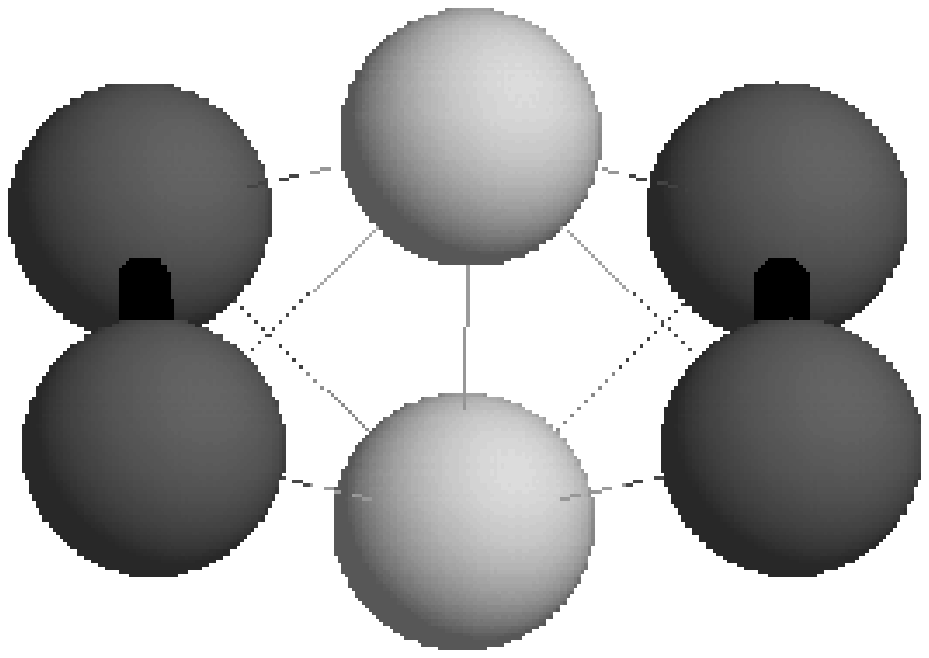,height=3cm}
\caption{Example of a 421-pair (paired atoms in light grey), with four common neighbors (dark grey), 
two bonds (black) and a longest chain of one bond.}
\label{421}
\end{figure}

Common neighbor analysis \cite{CNA93} (CNA) has been used to analyse cluster structures \cite{Uzi97,Uzi99}. 
CNA is a decomposition of the radial distribution function (RDF) according to the local environment of each
pair. We consider that the first peak in the RDF represents ``bonded" neighbors. As such, if $r_c$
is the first minimum in the RDF, we classify any pair separated by $r < r_c$
as a bonded pair. With this identification, any pair contributing to the RDF can be classified by a set
of three indices, $ijk$, which provide information on the local environment of the pair. The first
index, $i$, is the number of bonded neighbors common to both atoms. The second index, $j$, is the number
of bonds between this set of common neighbors. The third index, $k$, is the number of bonds in the longest
continuous chain formed by the $j$ bonds between common neighbors. Figure~\ref{421} shows a 421-pair
(light grey), with four common neighbors (dark grey), two bonds (black) and a longest chain of one bond.

\begin{figure}
 \hspace{1cm}
 \epsfig{file=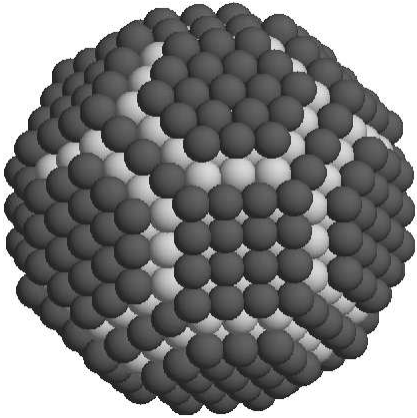,height=6cm}
\caption{A crystalline structure formed by relaxing a crystalline ball of 959 lead atoms. The darker atoms
denote surface atoms. Note the incomplete outer shell has plates of atoms on hexagonal (111)
faces. Hence we characterise this structure as an incomplete Wulff particle.} \label{r_fcc959}
\end{figure}

\begin{figure}
 \epsfig{file=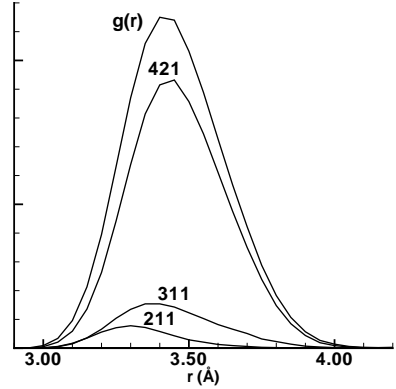,height=6cm}
\caption{Common neighbor decomposition of first peak in the RDF for a 959-atom cuboctahedron
(pictured in Figure~\ref{r_fcc959}) equilibrated at 200 K.} \label{cna_fcc200}
\end{figure}

Figure~\ref{r_fcc959} shows a 959-atom crystalline structure, formed by relaxing a ball of lead atoms 
``cut" from the fcc lattice. Noting that the incomplete outer shell has plates of atoms on hexagonal (111)
faces, we characterise this structure as an incomplete Wulff particle. The CNA decomposition of the 
first peak in the RDF for this cluster is shown in Figure~\ref{cna_fcc200}. Here, the RDF has been averaged 
over 100 time steps after the cluster was brought to thermal equilibrium at 200 K. The dominant contribution 
to the first peak comes from 421-pairs. This is characteristic of bulk fcc structure: each atom in a fcc crystal 
has twelve nearest neighbors and each of the pairs formed with these neighbors is a 421-pair. Also apparent
are 311-pairs, which predominantly arise from atoms on a (111) fcc face, and 211-pairs, which
predominantly arise from atoms on a (100) fcc face. One other pair type that contributes only weakly to
the first peak (and is thus difficult to distinguish in the decomposition shown in
Figure~\ref{cna_fcc200}) is the 200-pair, which is associated with atoms at the edges of the
incomplete outermost atomic layer of this cluster.

\begin{figure}
 \epsfig{file=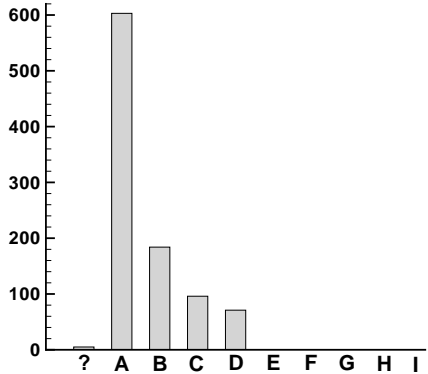,height=6cm}
 \caption{Classification of atoms in the 959-atom
incomplete Wulff particle (pictured in Figure~\ref{r_fcc959}) equilibrated at 200~K.} \label{sig_959_200}
\end{figure}

CNA is useful here because it allows one to distinguish between local atomic
arrangements, including fcc and icosahedral environments. In Table~\ref{CNA} we have listed the
classifications of CNA signatures used here to label the local environment of an atom (this 
classification is similar but not identical to that used by Cleveland, Luedtke and Landman\cite{Uzi99}). 
We note that
these signatures are based only on the CNA decomposition of the first peak in the RDF.
Figure~\ref{sig_959_200} shows the results of this classification scheme applied to the relaxed 
crystalline 959-atom structure of Figure~\ref{r_fcc959}. Here we see that predominantly fcc-bulk atoms with
(111)-face atoms, (100)-face atoms and (111)/(100) edge atoms are present.

\begin{table}
\caption{Description of CNA signatures used.}
\begin{tabular}{cl}
? & unclassified signature (possibly disordered)\\
A & fcc internal atom \\
B & fcc (111) face atom \\
C & fcc (100) face atom \\
D & fcc (111)/(100) edge atom \\
E & internal atom at a (111) fcc stacking fault \\
F & internal icosahedral atom (spine or central atom) \\
G & surface icosahedral apex atom \\
H & surface icosahedral (111)/(111) edge atom \\
I & ``shaved'' surface icosahedral (111)/(111) edge atom \\
\end{tabular}
\label{CNA}
\end{table}

\section{Binding energies of relaxed geometric structures}

Lim, Ong and Ercolessi \cite{LOE92} compared the energetics of the icosahedron sequence with the energetics
of the cuboctahedron sequence. These structures were relaxed to 0 K using the glue potential (\ref{glue})
and the total energies for each pair in this sequence were computed (recall that these polyhedra
have complete outer shells for 13, 55, 147, 309, 561, etc atoms). They found that the relaxed
cuboctahedra were strongly favoured energetically over the relaxed icosahedra.

One cannot directly compare the energetics of structures which do not have the same numbers of
atoms. In general, as cluster size increases, the binding energy per atom in the cluster also
increases, as the percentage of surface atoms decreases. This trend is often described by
the following relationship between binding energy per atom and cluster size \cite{LOE92}:
\begin{equation}
\label{fit}
E_b=A+B N^{-1/3} + C N^{-2/3}.
\end{equation}
By fitting $A$, $B$ and $C$ to the binding energies of families of structures, one can use
(\ref{fit}) to interpolate between clusters of complete outer shells and thus make comparisons between
families which do not occur in matching pairs.

In addition to the cuboctahedron and icosahedron sequences (looked at by Lim, Ong and Ercolessi
\cite{LOE92}), we also consider the sequences of the Ino decahedra\cite{ino69a} and Wulff cuboctahedra, making use of
(\ref{fit}) to compare energies by interpolation. The Wulff
construction minimizes the surface energy in a small crystal by taking into account the relative
surface energy of distinct crystal facets \cite{Herring51,Herring,Defay}. For this reason, one
expects Wulff clusters to be energetically favoured with respect to the alternative cuboctahedron
sequence. However, our results will show that any differences in binding energy are too small to be significant. We
also look at clusters extracted as a sphere from the fcc lattice \cite{Lewis97}. The number of
atoms in these ball clusters typically does not coincide with that of a closed-shell structure.
Nonetheless, when relaxed from the bulk they tend to resemble the Wulff form (see
Figure~\ref{r_fcc959}) as noted in the previous section; excess atoms in the incomplete outer shell contribute 
to hexagonal (111) facets.

\begin{figure}
\vspace{3mm}
 \epsfig{file=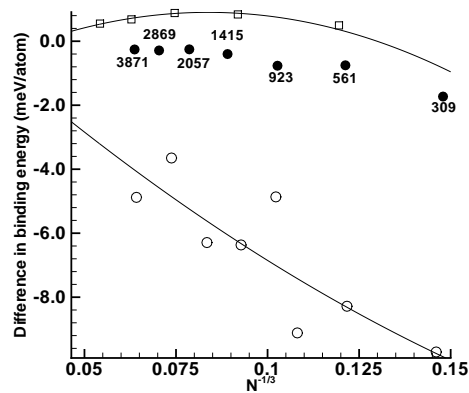,height=6cm}
\caption{Binding energies of relaxed crystalline clusters ($\Box$ Wulff particles, $\bullet$
cuboctahedra and $\circ$ incomplete shell Wulff particles) relative to the fit to the energy of the cuboctahedron sequence. For reference, the numbers of atoms of each member in the 
cuboctahedron sequence are shown.} \label{fcc_energy}
\end{figure}

The energies of the three types of crystalline model examined, are shown in Figure~\ref{fcc_energy}
together with the interpolations based on equation (\ref{fit}). For each, we have considered a
sequence of sizes, relaxing atomic configurations to 0~K and computing the binding energy per
atom. As might be expected, the incomplete Wulff particles were less favoured than the cuboctahedra or
Wulff particles, because of the incomplete (111) facets. However there appears to be little
distinction energetically between the Wulff and cuboctahedron structures. This contrasts with similar
comparisons in Lennard-Jones clusters \cite{Raoult89} and in nickel clusters \cite{Cleveland91},
both of which identify the Wulff particles as favorites.

\begin{figure}
 \vspace{3mm}
 \epsfig{file=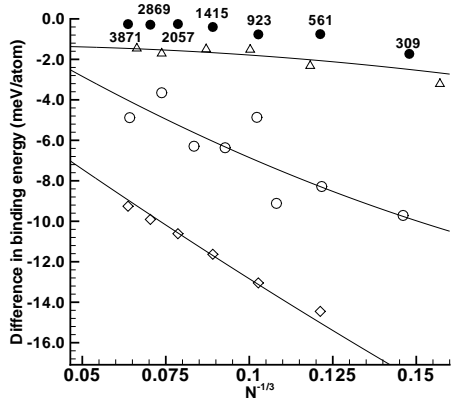,height=6cm}
\caption{Binding energies of relaxed clusters ($\bullet$ cuboctahedra, $\triangle$ Ino decahedra,
$\circ$ incomplete shell cuboctahedra, $\Diamond$ icosahedra ) relative to the fit to the energy of the cuboctahedron sequence. For reference, the numbers of atoms
of each member in sequence of cuboctahedron clusters are shown.} \label{energy}
\end{figure}

In Figure~\ref{energy}, we compare the energy of the MTP structures (decahedra and icosahedra) with
cuboctahedra and incomplete Wulff particles.
From Figure~\ref{energy}, we see that cuboctahedra are favoured, in general, over
MTP structures and fcc structures with an incomplete outer shell. Ino decahedra are also favoured
over the incomplete cuboctahedra, and all these structures are favoured over icosahedra. In
general, these results support the findings of Lim, Ong and Ercolessi \cite{LOE92}, that the
cuboctahedra are the most energetically favoured structures. Nonetheless, for certain cluster sizes
the relaxed Ino decahedra are the lowest energy structures considered. This is unexpected, because
the Ino decahedron is usually regarded as an unfavorable atomic arrangement; it is the more complex
re-entrant faceted model of Marks that is considered energetically competitive in small particles
\cite{Marks83,Marks84}. However, in the next section we see that none of these geometric models
constitute the lowest energy structures for a range of cluster sizes.

\section{Melting and freezing of clusters}

\begin{figure}
 \epsfig{file=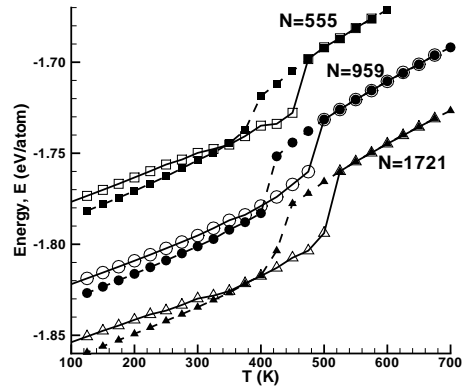,height=6cm}
\caption{Caloric curves for fcc ball clusters: 555-atom cluster, $\Box$ ; 959-atom cluster, $\circ$
; 1721-atom cluster, $\triangle$ . The empty symbols indicate heating and the filled symbols
represent cooling.} \label{melt}
\end{figure}

In a later study, Lim, Ong and Ercolessi \cite{LOE94} considered the simulated quenching of an 8217-atom liquid
cluster and observed the formation of an imperfect icosahedron upon solidification at constant energy. Here, we 
have performed a more general investigation of the melting and freezing of lead clusters.
Figure~\ref{melt} shows the caloric curves for several sizes of incomplete Wulff clusters (which were
first relaxed to their 0 K structures then heated). The caloric curves were constructed using a sequence of 
constant temperature simulations. The energy varies linearly with temperature on either side of the melting/freezing 
points, where a sharp change in energy occurs (latent heat of melting). The melting points of the clusters show a 
general increase with cluster size as expected, with all melting points lying below that of the bulk melting point 
(approximately 600 K). While ``hysteresis" in the melting/freezing point is typical in MD simulation, an effect 
of similar magnitude has been observed experimentally in lead clusters, where liquid lead clusters have been 
under-cooled by as much as 120~K without re-solidification \cite{Peters98}.

\begin{figure}
 \epsfig{file=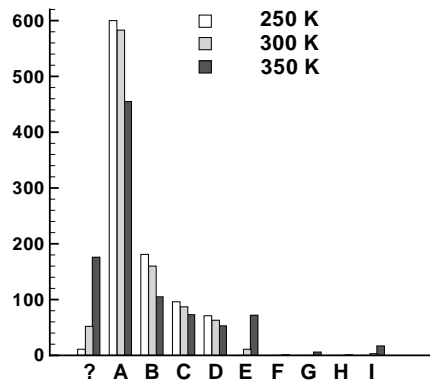,height=6cm}
\caption{Classification based on CNA signatures of atoms in the 959-atom cluster for
temperatures 250-350 K.} \label{sig_fcc}
\end{figure}

As clusters were heated towards the melting point, a number of interesting structural changes could
be identified from the CNA classification. Figure~\ref{sig_fcc} compares the CNA results of a
959-atom incomplete Wulff particle at temperatures 250-350~K, prior to melting. One can identify a general
trend towards disorder as the number of unclassified atoms (signature ?) increases. However, at
350~K we see a sharp increase in atoms with signature E, predominantly in the outermost {\em complete}
shell of the crystal. The signature E atoms arise as the incomplete (111) facets above these
atoms shift to form stacking faults at the surface. 

\begin{figure}
\epsfig{file=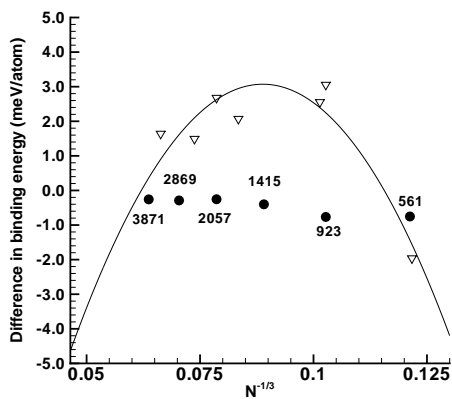,height=6cm}
\caption{Comparison of binding energies for cuboctahedra ($\bullet$)
and re-solidified icosahedron-like structures ($\bigtriangledown$) relative to the fit to the energy of the cuboctahedron sequence. The relative binding
energies of icosahedron-like clusters for sizes 555, 923, 959, 1721, 2057 and 3428 are shown here.
Apart from the 555-atom cluster, these structures are energetically favored over
cuboctahedra.} \label{g_energy}
\end{figure}

\begin{figure}
\hspace{1cm}
 \epsfig{file=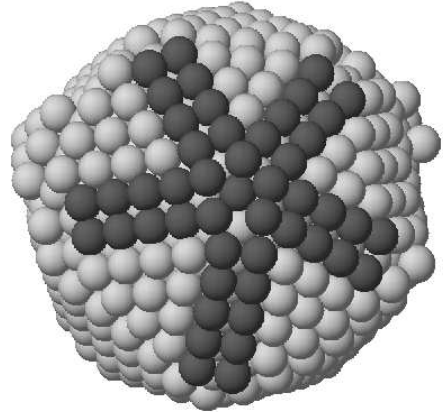,height=6cm}
 \caption{Re-solidified 959-atom cluster. Note
the axis of five-fold symmetry with ``shaved'' edges (darker atoms). Also visible are exposed excess 
atom layers on (111) faces.} \label{fcc959_f}
\end{figure}

\begin{figure}
\hspace{1cm}
\epsfig{file=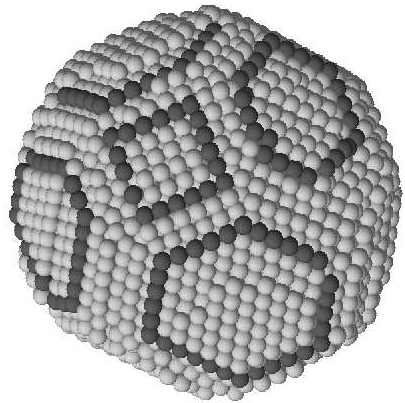,height=6cm}
\caption{Re-solidified 6525-atom cluster showing faceting typical of the Wulff particle, a cuboctahedron 
with hexagonal (111) faces.}
\label{cube}
\end{figure}

Returning again to the caloric curves (Figure~\ref{melt}), we see that, once re-solidified,
clusters have a lower energy than the original relaxed structures, indicating that atomic
rearrangement has probably occurred. For cluster sizes of between 555 and 3500 atoms, all the
re-solidified structures examined were found to be icosahedron-like. Figure~\ref{fcc959_f} shows
one example of a re-solidified 959-atom cluster. An axis of five-fold symmetry is clearly visible
at the surface and predominantly (111) faces are exposed. Figure~\ref{g_energy} compares the
binding energy of a number of re-solidified structures with cuboctahedra. As was apparent in
Figure~\ref{melt}, the re-solidified clusters appear to be favoured energetically over cuboctahedra
for a range of sizes from approximately 600 atoms up to about 4000 atoms. The interpolations
of the binding energy of each family, suggest that the fcc structures may become more favorable
above the 4000 atom size. One would expect to see a crossing of the two curves at large cluster
sizes, simply because the bulk structure should be favored in this limit. The largest re-solidified
structure in this study was the 6525-atom cluster shown in Figure~\ref{cube} (this was produced using 
the constant energy method detailed in section~\ref{structure}). Here, we clearly see an imperfect 
fcc Wulff particle structure emerging, which is consistent with these considerations (we consider the 
structure further in section~\ref{sec_resol_cub}).

Lim, Ong and Ercolessi \cite{LOE94} report a similar icosahedron-like structure that
formed upon re-solidification of an 8217-atom cluster (using the constant energy method
similar to the method detailed in the next section). From considerations of the binding energies 
of geometric icosahedra and cuboctahedra (see Figure~\ref{energy}), they expected the quenched 
liquid droplet to form a fcc cuboctahedron. The appearance of an imperfect icosahedron was attributed 
to an insufficient relaxation time. While the interpolation (\ref{fit}) indicates that for clusters of 
sizes of up to 4000 atoms, the icosahedron-like structures are favoured over fcc, extrapolating (\ref{fit})  
suggests that clusters above 4000 atoms favor crystalline structures. We conclude that the failure of 
earlier studies to find fcc structures on re-solidification is in part due to shorter relaxation times. 
We note that the relaxation times used here, are several times longer than those of Lim, Ong and 
Ercolessi \cite{LOE94}.

The re-solidified icosahedra in this section were produced using constant temperature simulations. In the following
section, we show that these icosahedra are also produced from liquid droplets by freezing at constant energy. We 
will also characterise the structure of these icosahedra using CNA analysis.

\section{Re-solidified icosahedra}
\label{sec_resol_ico}

\subsection{Structure of re-solidified icosahedra}
\label{structure}

In this section, we examine the structure of a ``shaved'' re-solidified icosahedron cluster produced
using constant-energy re-solidification. Starting with a perturbed 2057-atom Wulff particle (each atom 
was perturbed in a random direction with mean displacement 0.1 \AA), the cluster was melted and then 
brought to equilibrium at 900 K for approximately 1 ns. The cluster was then quenched rapidly below 
the melting point, where it was allowed to equilibrate at constant energy for approximately 10 ns. The 
structure was then slowly relaxed to 0 K where the binding energy was computed 
and the structure was determined with CNA analysis. The re-solidification procedure is similar to that 
used by Lim, Ong and Ercolessi \cite{LOE94}.

\begin{figure}
 \epsfig{file=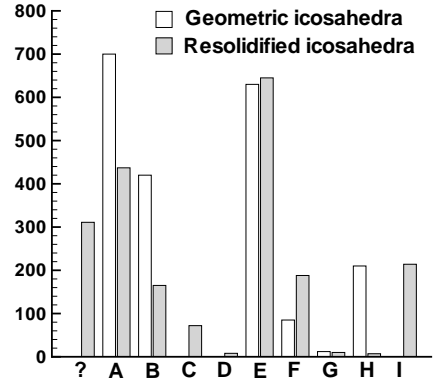,height=6cm}
 \caption{The CNA classification of a 2057-atom icosahedron
compared with the classification of a 2057-atom re-solidified cluster.} \label{sig_ico}
\end{figure}

Figure~\ref{sig_ico} compares the CNA classification of a 2057-atom icosahedron, before melting, with
a re-solidified 2057-atom cluster. We see that disorder seems to have increased with
re-solidification (signature ?), and the number of internal fcc atoms (signature A) and (111) face
atoms (signature B) have also decreased. Also of note is the appearance of (100)-faceting
(signature C) as well as the removal of edge atoms along a twin plane (leading to signature I
atoms). An increase in the number of hcp-type (signature E) atoms is also apparent. 

This re-solidified icosahedron is a surface-reconstructed, or ``shaved", icosahedron, where the 
(111)/(111) facet edges and the icosahedral apex atoms on the outer shell have been redistributed 
in the cluster, and the (111) layers have moved to form stacking faults with respect to the internal fcc 
tetrahedra of the icosahedron (leading to an increase in signature E atoms). While the disappearance 
of the interior fcc atoms appears, in part, to be due an increase in disorder, the shaving of (111)/(111) 
edges gives the internal fcc atoms below these edges icosahedral coordination, resulting in these atoms 
being classified as ``extra'' internal icosahedral atoms (signature F). This ``shaved'' icosahedron surface 
is the same surface reconstruction as that noted by Doye and Wales in their investigation of icosahedra 
using a Morse potential \cite{Doye97}. Their study found that geometric icosahedra underwent thermally-induced 
surface reconstructions prior to melting to form the same ``shaved'' icosahedra seen here.

The average binding energies of the atoms according to classification is given in Table~\ref{shaved}.
The replacement of the (111)/(111) facet edges, with an average binding energy of $1.463$ eV/atom, by
shaved facet edges with an average binding energy of $1.631$ eV/atom, considerably increases the total
binding energy. This compensates for the increase in internal icosahedral atoms (signature F, $1.997$
eV/atom) at the expense of internal fcc atoms (signature A, $2.016$ eV/atom). This rearrangement of 
surface atoms, with an icosahedral internal structure, is energetically favoured over the crystalline 
cuboctahedra for cluster sizes from 600 to 4000 atoms (Figure~\ref{g_energy}). In their original study 
of lead clusters using the potential (\ref{glue}), Lim, Ong and Ercolessi \cite{LOE92} concluded that 
fcc crystalline structures were energetically favoured over this range. However, they did not consider 
the effect of the ``shaved'' surface reconstruction which, as is evident here, stabilises the icosahedra.

\begin{table}
\caption{Comparison of average binding energies of atoms in an icosahedron prior to melting and
after re-solidification.}
\begin{tabular}{c|c|c|c|c}
 & \multicolumn{4}{c}{Cluster} \\ \hline
& \multicolumn{2}{c}{Closed-shell} & \multicolumn{2}{c}{Re-solidified} \\ Signature & N & E & N & E
\\ & & (eV/atom) & & (eV/atom) \\ \hline & & & & \\ ? surface & - & - & 169 & 1.541 \\ ? interior &
- & - & 142 & 1.975 \\ A & 700 & 2.016 & 437 & 2.007 \\ B & 420 & 1.625 & 165 & 1.703 \\ C & -  & -
& 72 & 1.638 \\ D & - & - & 8 & 1.471 \\ E & 630 & 2.010 & 645 & 2.010 \\ F & 85 & 1.992 & 188 &
1.997 \\ G & 12 & 1.056 & 10 & 1.674 \\ H & 210 & 1.463 & 7 & 1.533 \\ I &  - & - & 214 & 1.631 \\
\end{tabular}
\label{shaved}
\end{table}
 
\subsection{Re-solidification of liquid droplets}

\begin{figure}
 \epsfig{file=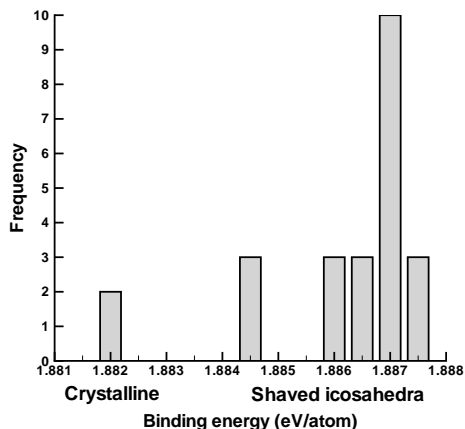,height=6cm}
\caption{Distribution of binding energies for 25 re-solidified 2057 atom clusters. Two re-solidified
clusters were identified as Wulff particles (energies of 1.882 eV/atom) while the remaining 23 were identified
as ``shaved'' icosahedra (energies above 1.884 eV/atom).}
\label{energy-dist}
\end{figure}

In this section we have examined a total of twenty-five 2057-atom cluster structures produced by 
re-solidification at constant-energy using the scheme outlined in the previous section. This enables us to 
investigate the reproducibility of the ``shaved'' icosahedron structures. The distribution of binding 
energies for the twenty-five re-solidified clusters is shown in Figure~\ref{energy-dist}. Two re-solidified 
clusters were characterised by CNA analysis as Wulff particles, with energies of 1.882 eV/atom (equal to the binding 
energy for a perfect geometric Wulff particle). The remaining twenty-three were ``shaved'' icosahedra 
(energies above 1.884 eV/atom). The two ``shaved'' icosahedra with energy 1.884-1.885 eV/atom 
had surface defects (one edge not shaved). However, the majority of the ``shaved'' icosahedra are clustered 
about an average energy of 1.887 eV/atom. Thus we can see that with equilibration times of 10 ns kinetic 
trapping in less optimal structures is uncommon.     

The repeated occurrence of the ``shaved'' icosahedron structure in these trials suggests that it is the preferred 
structure at this size for the potential (\ref{glue}). This method did not produce any structures which were 
energetically more favoured than the ``shaved'' icosahedra in any of the trials. Furthermore, these ``shaved'' 
icosahedra appear upon re-solidification using both constant temperature (section IV) and constant energy 
(section V) simulations at a variety of temperatures, cooling rates and equilibration times (up to the 10 ns 
used in this section). In the next section, we see that by heating geometric icosahedra, a transformation
to ``shaved'' icosahedra occurs prior to melting.  

\subsection{Structural instability of geometric icosahedra}
\label{instab}

\begin{figure}
 \epsfig{file=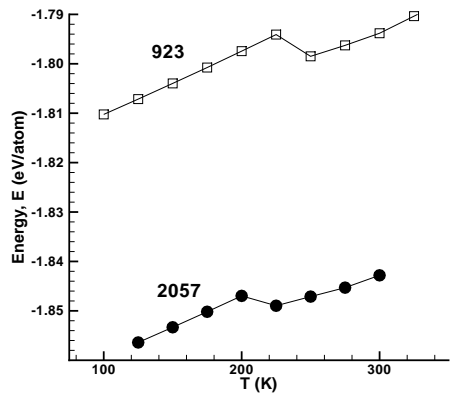,height=6cm}
\caption{Caloric curve for 923 ($\Box$) and 2057 ($\bullet$) atom icosahedron clusters. The 923-atom
cluster undergoes a structural transition at 250~K. The 2057-atom cluster undergoes a structural
transition at approximately 225~K.}
\label{ico_str}
\end{figure}

In this section we show that geometric icosahedra can be induced by heating (below the melting point) 
to form ``shaved'' icosahedra in the same manner as that seen by Doye and Wales \cite{Doye97}.
Figure~\ref{ico_str} shows the caloric curve for two icosahedron clusters 
of 923 and 2057 atoms. These clusters have been prepared by first relaxing a geometric icosahedron to 0~K. 
The clusters were then brought to equilibrium at 100~K, and then heated in steps of 25~K to 300~K.
At each temperature, the clusters were allowed to equilibrate over $5 \times 10^5$ time steps at 
constant temperature. Between 200~K and 250~ K, the energy of clusters is seen to drop sharply, 
suggesting a change in cluster structure.
The new structure that appears at temperatures above this transition point is essentially the same
as the ``shaved'' icosahedron clusters obtained in the previous section after re-solidification.
We see that geometric icosahedra relax to the equilibrium configuration (the ``shaved'' 
icosahedron) once heated to a certain temperature. The temperature at which this relaxation occurs 
is seen to depend on cluster size (which determines the energy barrier for this process) but will 
also depend on the time scale of the simulation. 

\section{Re-solidified Wulff particle}
\label{sec_resol_cub}

Here we discuss the structure of the re-solidified 6525-atom crystalline structure (shown in
Figure~\ref{cube}) produced by freezing a 6525-atom icosahedron using the constant energy
method outlined in section~\ref{structure}. Apparent in Figure~\ref{cube} is the appearance of 
hexagonal (111) faces. This is interesting because a 6525-atom cluster can form a perfect, closed-shell, 
cuboctahedron with triangular faces but not a perfect Wulff particle (cuboctahedron with hexagonal faces). 
The average binding energy per atom of this re-solidified cluster is 1.92953 eV, as opposed to the binding energy 
of 1.92982 eV for a model 6525-atom cuboctahedron with triangular (111) faces. The difference of 0.3 meV per
atom, in favour of the cuboctahedron with triangular (111) faces, does not explain the preference
for the Wulff form. However, it is probable that the Wulff particle represents a local minimum in
the energy in configuration space and that the liquid droplet is closer in configuration space, due
to spherical symmetry, to this form of cuboctahedron. A lengthier relaxation period upon
re-solidification may see the cluster form a cuboctahedron with triangular (111) faces.

\begin{figure}
 \epsfig{file=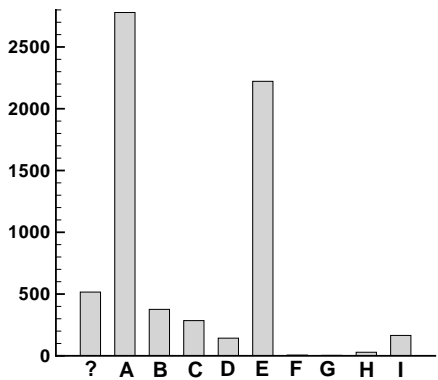,height=6cm}
\caption{The CNA classification of a 6525-atom re-solidified
cluster.} \label{sig_6525}
\end{figure}

\begin{figure}
 \epsfig{file=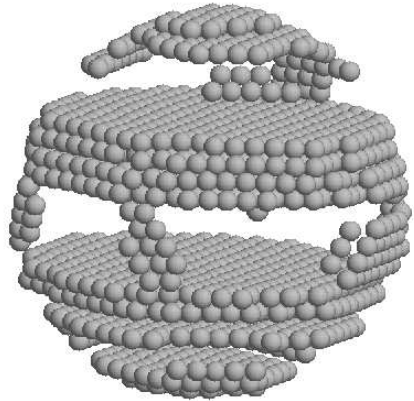,height=6cm}
 \caption{Atoms which show signature E (stacking fault or twinning plane) in
the re-solidified 6525-atom cluster.} \label{fault}
\end{figure}

The CNA classification of the re-solidified cluster is given in Figure~\ref{sig_6525}.
The profile is essentially fcc, except for the large number of type-E atoms. These arise from
internal stacking-fault defects in the structure, either when a twin plane occurs (the fcc (111)
packing sequence goes, e.g.: ...ABC\underline{A}CBAC...), or when a skip (deformation fault) in the
sequence occurs (e.g.: ...ABC\underline{B}CABC...). Note that the stacking fault planes are all
oriented in the same direction. The signature E classification (Table~\ref{CNA}) is based on the
decomposition of the first peak of the RDF, and does not distinguish between the two types of
defect (although it would be useful to do so \cite{Warren}). The E-type fault planes are shown in
Figure~\ref{fault}. Ignoring the cap regions, there is a set of four consecutive fault planes in
the upper half of the cluster which is a series of deformation faults that has effectively produced
a narrow band of hexagonal-close-packed structure. In the lower half of the cluster, a pair of 
deformation fault planes occur, and below them is a single twin-plane fault.

\section{Diffraction patterns}
Diffraction is a valuable experimental probe of nanoparticle structure \cite{Zanchet00}. The
diffraction pattern of a nanoparticle sample is a ``powder" pattern with radially symmetric
intensity that can readily be calculated from the radial distribution function, using the Debye
equation of kinematic diffraction theory \cite{Guinier}. The so-called ``interference function'' is
closely related to the distribution of intensity in an experimental diffraction pattern and is
given by

\begin{equation}
J(s) = \frac{1}{N}\sum_{n, m} \frac{\sin (2 \pi s r_{mn})}{2 \pi s r_{mn}} \; , \label{eq_d}
 \label{eq_debye}
\end{equation}
where the scattering parameter $s = 2\sin (\theta) / \lambda$, with $\theta$ equal to half the
scattering angle and $\lambda$ the wavelength of the radiation. $N$ is the number of atoms in the
cluster and $r_{mn}$ is the inter-atomic distance between atoms $m$ and $n$. $J(s)$ is
dimensionless; it is related to the distribution scattered intensity, but is independent of the
type of radiation and, for monoatomic clusters, the atomic scattering factor \cite{Guinier}.

It is generally difficult to interpret diffraction measurements on nanoparticles because of the 
likely occurrence of non-crystalline and imperfect structures, as well as sample size distribution 
and inhomogeneity, so the standard techniques of x-ray crystallography do not apply.
Determination of structure is usually based on the careful comparison of experimental data with 
diffraction patterns calculated from models of particle structures. When a close correspondence 
between observations and model-based profiles is obtained, the principal characteristics of the 
model structure can generally be inferred to exist in the actual cluster sample. The best known 
and most successful example of this approach is probably the electron diffraction studies of 
rare-gas clusters that were interpreted using MD models \cite{Farges83,Farges86}.

\begin{figure}
 \epsfig{file=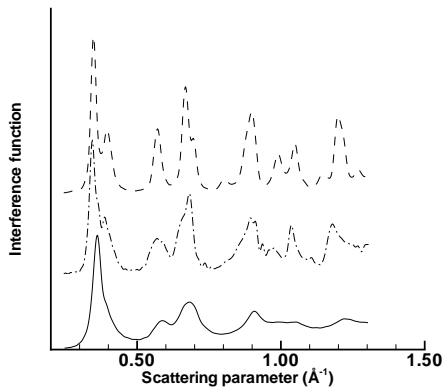,height=8cm}
\caption{Interference function for the 3428 atom re-solidified cluster (---) compared to that for
the original decahedron (- $\cdot$ -) and a 2869 atom icosahedron (- - -). } \label{fig_dif_3428}
\end{figure}

\begin{figure}
\epsfig{file=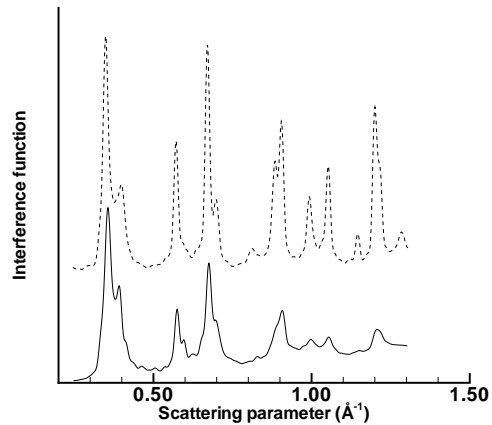,height=8cm}
\caption{Interference function for the 6525 atom re-solidified
cluster (---) compared to that of an ensemble of twinned structures (- - -).} \label{fig_dif_6525}
\end{figure}

Typically in experimental studies, geometric models are used to provide the needed 
candidate diffraction profiles. Thus it is of interest to consider how MD models of structure compare with 
geometrical models. With this in mind, we consider the interference functions of the two larger 
clusters that have been generated by re-solidification in this study. Figure~\ref{fig_dif_3428} 
shows the pattern obtained from the 3428-atom cluster, and Figure~\ref{fig_dif_6525} shows the 
pattern for the 6525-atom cluster.

The 3428-atom cluster was obtained after melting a decahedron, so in Figure~\ref{fig_dif_3428} we
also show the interference pattern of a 9-shell decahedron (3428 atoms -- the initial structure)
and a 9-shell icosahedron (2869 atoms). Several comments are in order. The MD simulated cluster has
a smoother interference function that either of the geometric models; the features are broader and
the MD profile is also noticeably attenuated with increasing $s$. This is all consistent with a
high degree of static disorder in the MD structure. The four significant peaked features in the MD
curve (at: $s\simeq$ 0.37, 0.6, 0.7, 0.9) are more closely matched by the profile of the
icosahedron, although the decahedron also has peaks at these positions. In either case, the
position of the features in the MD profile is shifted to higher $s$ (the shift is about $0.01\,{\rm
\AA}^{-1}$), indicating a contraction of the length scale for the MD structure  of the order of
3~\%, compared to the geometric forms.

Clearly, from the viewpoint of a diffraction measurements, our MD cluster would be better matched
by icosahedral models than either fcc structures or decahedra. Diffraction therefore supports CNA
in this case, although they are fundamentally different types of analysis: CNA characterises the
local environment of atom pairs, whereas diffraction is a reciprocal space view of the radial
distribution function, and hence tends to identify longer range correlations in a structure.

The 6525-atom cluster has been identified by CNA as having the fcc structure of the bulk.
Nevertheless, Figure~\ref{fig_dif_6525} shows, again, the attenuation of the structure's
interference function with increasing $s$, indicating substantial disorder. Figure
\ref{fig_dif_6525} also shows an interference function calculated for an ensemble of perfect
twinned structures. These were fcc ball clusters with random stacking faults along the (111) axis.
A set of 25 interference functions from such models were averaged to produce the profile shown. The
main features of the MD structure interference function coincide with the model ensemble pattern,
although the relative intensities differ substantially. It is also interesting that the location of
features in the MD profile and the static models profile are not shifted, as they were for the
3428-atom structure. This means that the lattice parameter predicted by the MD simulation agrees
with the bulk value used to construct the models.

\section{Discussion}

We have considered the binding energies of relaxed lead clusters for a range of particle sizes and
cluster structures using the ``glue'' potential (\ref{glue}). For medium cluster sizes (600 to 4000 atoms), we found 
that a surface-reconstructed
icosahedron structure was the lowest energy structure of those considered. These are shaved
icosahedra, with atoms removed from the edges and missing five-fold apex atoms, and surface atoms 
relocated to form a stacking fault with respect to the internal fcc tetrahedra of the icosahedron. Earlier 
theoretical studies of the energetics of MTPs by Marks \cite{Marks84}, suggested that the removal of
icosahedral apex atoms would lead to improvements in energy. This was supported by atomistic energy
calculations for Lennard-Jones clusters \cite{Raoult89}, but not in the case of nickel
\cite{Cleveland91}. In practice, it would be virtually impossible to observe this shaving by microscopy, 
as explained by Buffat, Fl\"ueli, Spycher, Stadelman and Borel \cite{Buffat91}, 
and we are not aware of any reported observations. However, when indeed microscopy can identify cluster 
structure as icosahedral, particles generally appear to be imperfect \cite{Buffat91}. This raises the 
interesting possibility that the MD structures grown here may compare well with typical high-resolution
images of some small particles.

In this work we did not identify signs of icosahedral precursors to melting as reported in simulations of
gold clusters\cite{Uzi98}. These, however, might be revealed by constant energy simulations (as used in the 
studies of gold\cite{Uzi98}) of fcc clusters near the melting point. We speculate here that these precursors, 
if found, will tend to resemble our shaved icosahedral structure more closely than the geometric icosahedra. 
Indeed, in section~\ref{instab}, we saw that by heating, geometric icosahedra could be induced to undergo a 
surface reconstruction to the shaved icosahedra prior to melting.

Our finding that icosahedron, rather than fcc, structures are preferred for certain sizes contrasts
with an earlier MD study of lead that used exactly the same potential model, but with shorter
simulation times \cite{LOE92},\cite{LOE94}. While at the time of that study there were no experimental reports
of lead icosahedra, a recent investigation has indicated that they do occur abundantly
\cite{Hyslop00}. Electron diffraction experiments on a beam unsupported lead clusters has
identified both decahedral and icosahedral signatures in the diffraction patterns for beams with
average cluster size ranging between approximately 3 and 6~nm. Preliminary analysis of those 
results has relied on geometric models of cluster structures, however the icosahedral component in
the data is strong and unambiguous. On the other hand, the decahedral component, identified in
larger clusters, may also be compatible with the kind of imperfect and faulted structure that has
been obtained in the present study for the 6525-atom cluster.

The imperfect structure of this 6525-atom cluster is interesting from the perspective of
experimental observations. The cluster's main feature is a series of (111) stacking faults, all
parallel to the same axis. These are mainly deformation faults, although one twin plane is present.
Very recently, de Feraudy and Torchet commented that parallel stacking faults were most unlikely to
occur in unsupported argon clusters \cite{DeFeraudy00}. Our findings suggest that, for lead, this
may not be the case. If so, this poses yet more difficulty in interpreting cluster
diffraction measurements. De Feraudy and Torchet point out that if faulted clusters of this type
are to be considered, then crystallographic-based calculations based on a faulted bulk crystal
model are not satisfactory: detailed atomistic modeling is required \cite{DeFeraudy00}. However, our
simple attempt to produce approximate atomistic models of clusters containing parallel fault planes
have been unsatisfactory. As shown in Figure \ref{fig_dif_6525}, the degree of accord with the MD
structure is quite poor, and would not be accepted in an objective analysis of real experimental
data.

In conclusion, MD simulations of lead clusters have identified new non-crystalline low energy
structures after re-solidification of medium-sized liquid droplets. These forms are
surface reconstructed icosahedra; their binding energies are higher than the corresponding perfect
(relaxed) icosahedron as well as the Wulff form. We have also shown that sufficiently large liquid
droplets re-solidify into imperfect crystalline structures, resembling the Wulff particle, provided
adequate equilibration times are used. We have made a preliminary investigation of the effects of
temperature on structure, noting the appearance of stacking faults at the surface of incomplete
cuboctahedra and a similar solid-to-solid structural transition in geometric icosahedra.

\section*{Acknowledgments}
We would like to acknowledge the use of the Ames Laboratory Classical MD code. We are also grateful
to F. Ercolessi for making the lead ``glue'' potential freely available. We would also like to thank
Jonathon Doye for his comments.

\end{multicols}

\end{document}